\documentclass[11pt]{article}
\usepackage{graphics,latexsym,amsfonts}
\usepackage{graphicx}
\usepackage{float}
\usepackage{picture}
\usepackage[english]{babel}
\usepackage{amsmath,mathtools,amsthm}
\usepackage{enumitem}
\usepackage{color}
\usepackage{comment}
\usepackage{lineno}
\usepackage{algorithm}
\usepackage{algpseudocode}
\usepackage{graphicx}
\usepackage{caption}
\author{
A.\ E.\ Biondo\thanks{Department \ of Economics and Business,
University\ of Catania, Italy; ae.biondo@unict.it}$\:$,
A.\ Pluchino\thanks{Department\ of Physics and Astronomy,
University\ of Catania and INFN Sezione di Catania, Italy; alessandro.pluchino@ct.infn.it}$\:$,
A.\ Rapisarda\thanks{Department\ of Physics and Astronomy,
University\ of Catania and INFN Sezione di Catania, Italy; Ccomplexity Science Hub Vienna; andrea.rapisarda@ct.infn.it}}

\title{\bf Modelling Surveys Effects in Political Competitions} 

\begin{document}
\date{ }
\maketitle

\begin{abstract}
\noindent In this paper we study the impact of news media and public surveys  on the electoral campaigns for  political competitions. We present an agent-based model that addresses the effective influence  of surveys in orienting the opinions of voters before elections. The  dynamics of electoral consensus is studied as a function of time by investigating different possible scenarios and the effect of periodic  surveys, on the opinions of a  small community of voters represented as a network of agents connected by realistic social relationships. Our simulations  show the possibility to manage  opinion consensus  in electoral competitions. 
    
\end{abstract}
\section{Introduction}

The relation between the news media and the electoral competition has attracted growing attention in literature quite recently, as shown in \cite{gerber2009does}. The way in which the information is provided to the public may reflect a position with regard to parties or politicians. Thus, for example, \cite{groseclose2005measure} propose a measure of political orientation by locating different media outlets on the political spectrum on the basis of the similarity of the experts used by the media outlet and those cited by members of Congress. There exist a relevant number of studies reporting correlations between media usage and reported behavior as, among others, \cite{clarke1978newspapers, miller1979type, bybee1981mass, garramone1986mass, lieske1989political, brians1996campaign, dalton1998partisan, hibbing1998media}. 

A more recent paper by \cite{kull2003misperceptions}, compared the perception of the Iraq war of people who viewed Fox News with those who did not, in order to show that the habit to follow some informative channel can affect the perceptions of reported news. This raises a point, for people have the tendency to seek out information that agrees with their pre-existing views. Such a conslusion has been firstly documented long time ago, in \cite{brock1965commitment} and then in \cite{sweeney1984selective}. However, more recently, some theoretical work on the hypothesis that individuals adapt their choices about media sources according to the similitude with their political perspective have been conduced by \cite{mullainathan2005market, gentzkow2006media}. Cited evidence shows that a sort of perverse reinforcing mechanism operates without helping any improvement of knowledge and, instead, increasing the political discussion, as shown by \cite{mondak1995media, mondak1995newspapers}. 

Such a phenomenon has lead the recent literature on social communication to investigate new methodologies, based on the concept of echo-chambers, in which a spontaneous clustering emerges among people, as the result of a self-reinforcing preferential attachment dynamics. This process has been validly replicated on the internet, where both the chances of contacts among people and the information sharing are increasingly relevant. Following the idea of ``public sphere'' provided by \cite{dahlgren2005internet}, i.e. the communicative environment in which it is possible to circulate ideas and information, some authors argue that the internet usage has boosted a passive people's exposure to the political debate, as shown for example, in \cite{wojcieszak2009online} and in \cite{brundidge2010encountering}. Some other authors suggest, instead, that a selective mechanism operates for people surfing the web in a selective manner, according to their \textit{a priori} political views. In this case, then, an active choice would operate in seeking the most appreciated information, as held for example, by \cite{bimber2003campaigning,kushin2009getting,stroud2010polarization}. This debate is not being put forward here: what matters, for our purposes, is the evidence that news media affect electoral participation of citizens, as shown by \cite{stromberg2004mass, oberholzer2009media, snyder2010press, gentzkow2011effect, schulhofer2013newspapers, drago2014meet}, among others. For example, as explained in \cite{colleoni2014echo}, social networks may help in recognizing the political orientation on the basis of the shared contents. 

The strict relationship between informative campaigns, surveys and news management on one side and the political orientation of the public on the other, operates thus similarly to advertising for consumption activities. There emerges the chance to finely tune it, by means of a series of stimuli induced to manage consensus. For example, it can be strategic to know to which extent news media can affect the chances of incumbent politicians to be elected again, or whether specific information provided can create competitive advantage for a type of politicians. The debate on both points is still open. Empirical evidence does not support unambiguously a direct or inverse causation effects, as shown in contributions by \cite{prat2011political}, \cite{ansolabehere2006television, gentzkow2011effect, snyder2010press, drago2014meet}. However, an increase in provided information (in terms of number of news media) is shown to reduce the advantage of incumbents and, thus, to increase both the turnover and the quality of politicians, as in \cite{besley2006handcuffs}. This should also reduce the chance to cover corruption, even if \cite{gentzkow2006television} and in \cite{george2008national} show also the evidence that an increase in the supply of news media may lead to a negative effect on electoral participation due to a crowding-out effect on the existing type of news media. 

A more delicate point is shown by several authors: provided information is often far from being objective, e.g., \cite{goldberg2014bias, alterman2003liberal, bagdikian2004new, davies2008flat}. More explicitly, \cite{posner2005bad} specifies that media lie about the news, by choosing what to say and what to hide (see \cite{anderson2012media}), by selecting the timing of the news diffusion (see \cite{larcinese2011partisan}), by creating the context in which the information can implicitly suggest the desired reaction, apparently spontaneous (see \cite{gentzkow2010drives}). In \cite{sobbrio2011citizen, sobbrio2011indirect} it is addressed directly the existence of a media bias, which descends from the way journalists gather information from their sources. Such a dramatic result may derive from choices made by journalists themselves or their media owners, as explained respectively in \cite{baron2005competing} and in \cite{anderson2012media}, but also from eternal pressures exerted on the media by politicians, as argued in \cite{besley2006handcuffs}, lobbies, as in \cite{baron2005competing, sobbrio2011citizen, sobbrio2011indirect, petrova2012mass}, or advertisers, as in \cite{ellman2009papers, blasco2011paying, germano2013concentration}.

The main motivation of this paper is to show, by means of agent-based simulations, whether and to which extent, the informative signals can effectively play a role in political competitions among participating parties or coalitions. During these last years, agent-based models have been extensively adopted in order to investigate emergent behavior and describe the implications of complexity in several socio-economic phenomena as in \cite{biondoorder, biondo2012return, biondo2013beneficial, biondo2013random, biondo2013reducing, biondo2014micro, biondo2015modeling, biondo2016multi, biondo2016order, biondo2016perfect, biondo2017informative, biondo2017multilayer, pluchino2010peter, pluchino2011accidental, pluchino2011efficient} among many others.

Several scenarios will be presented to capture the influence of  surveys - i.e. those statistical investigations, based more or less explicitly on interviews, which reports the stated preference a sample of voters during an electoral campaign. We propose a model which can rely on the complexity of interactions among members of a small community, in order to refer to the relevance of the effect of the media bias. These aspects have been only partially investigated. In particular they were empirically studied in  the paper of \cite{dellavigna2007fox}, who analyzed the consequences of the diffusion of Fox News in several US towns between $1996$ and $2000$ to show that between $3\%$ and $\%8$ of Fox News' viewers where induced to vote Republican. Other examples of such a relevant branch of literature are, among others: \cite{chiang2011media} on the effects of US newspapers endorsement of presidential candidates on voter behavior; \cite{enikolopov2011media} on the variation in the reception of the sole Russian TV independent channel ``NTV'' to study the impact on the vote share of government and opposition parties; \cite{dellavigna2011unintended} on the evidence showing that the reception of nationalistic Serbian radios signal increases the vote share of extremist nationalistic parties in the neighboring Croatian region; \cite{durante2012partisan} on the effects on public television news programs in Italy after the electoral victory of the coalition lead by Berlusconi in 2001. 

The possibility to find some statistical regularities in the dynamics of the electoral campaigns descending from a suitable management of news media is appealing for parties, but not only. The political orientation of voters must be free and self-determined. The possibility that external stimuli may play a hidden role is dangerous and should be carefully studied and eventually regulated for a conscious and responsible use of the democratic mechanisms. The paper is organized as  follows: in section two, the model is presented; section three contains simulation results and discussion; section four presents conclusive remarks.

\section{The model}

\begin{figure}
\begin{center}
\includegraphics[width=2.8in,angle=0]{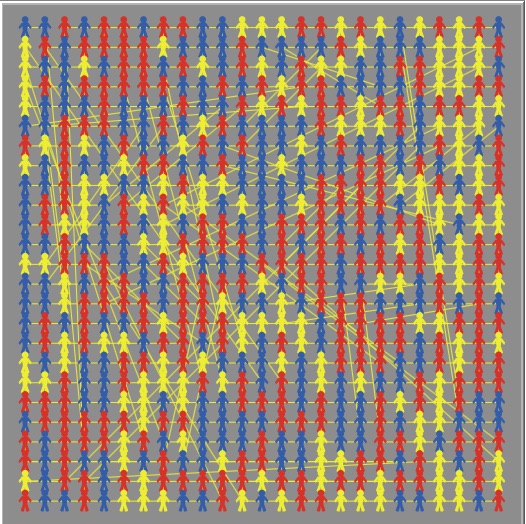}
\caption{
An example of community represented by a small-world $2D$ lattice with $N=298$ nodes. In the simulations we considered greater values for $N$. The different colors of the nodes (red, blue, yellow) indicate the different opinions of the agents and correspond, respectively, to individuals either voting Red party or voting Blue party or non-voting at all (undecided).  
}
\label{fig1} 
\end{center}
\end{figure}

The  goal of the present study is to show the effects of repeated and periodic public surveys on the voting orientation of a relatively small community of people during a time interval  of several months before a political election. In this respect,  we adopt a "canonical ensemble" perspective, i.e. the test community we consider is {\it only exposed} to the surveys' results, which report - through different media channels - the voting intentions of a much larger population, and can be only influenced by, but cannot influence, them. In this respect  you can think, for example, to the residents of a small village, or to a small Facebook community, which are periodically informed by media about the general national political orientation.    
\\
\\
\textit{Network description}
\\
The community we have in mind can be  modeled as an undirected small-world network with $N$ nodes, where each node is an individual (agent), $A_i$ ($i=1,...,N$), able to share information with, on average, four neighbors, some of them linked with long range edges (see Fig.\ref{fig1}). Such a network topology, obtained from a regular $2D$ lattice, by means of a rewiring procedure with probability $p_r=0.02$, ensures that the information flows quickly propagate through weak ties through the system and reaches also agents far away from each other in terms of degree of separations. At the same time, the existence of strong ties preserves necessary clustering properties that characterize real social networks. 
\\
\\
\\
\textit{Agents description}
\\
Each node (i.e. each agent) of the network has a color indicating the political preference of that agent for one of two parties, the Red party (red nodes) and the Blue party (blue nodes). Further, a Yellow color characterizes an undecided agent whose preference is not oriented towards any of the two parties and, therefore, corresponds to the non-voting intention at the elections (actually, the Yellow color may indicate all the different positions - indifference, scarce interest, confusion - which cannot be translated into a vote intention). Capital letters $R$, $B$ and $Y$ indicate the three possible choices, corresponding to the described preferences and thus we define a status variable $O_i$ ($i=1,...,N$) assuming one of these values, for each agent. We also indicate with $N_R$, $N_B$ and $N_Y$ the size (number of agents) of the three resulting groups, so that $N_R+N_B+N_Y=N$. 

%%%%%%%%%%%%%%%%%%%%%%%%%%%%%%%%%%%%%%%%%%%%%%%%%%%%%%%%%%%%%%%%%%%%%%%%
%%%%%%%%%  FIG. 1
%%%%%%%%%%%%%%%%%%%%%%%%%%%%%%%%%%%%%%%%%%%%%%%%%%%%%%%%%%%%%%%%%%%%%%%%%%

In our model each agent is characterized by two real variables, $IR_i$ and $IB_i$ ($i=1,...,N$), which represent his/her own intensity of believing into, respectively, Red party or Blue party. At the beginning of a simulation ($t=0$), for agents who belong to one of the two parties, one of these variables assumes a value selected, with uniform probability, in the interval $[I_{min},I_{max}]$ (with $I_{min}, I_{max} \in [0,1]$ and $I_{min} < I_{max}$), while the other variable is set to zero. Of course, an agent $A_k$ belonging to Red party ($O_k=R$) will have $IR_k>0$ and $IB_k=0$; vice-versa, an agent belonging to Blue party ($O_k=B$) will have $IB_k>0$ and $IR_k=0$. In both these cases, depending on the value of $IB_k$ or $IR_k$, the same agent could be also defined a strong believer (for values close to 1) or a weak believer (for values close to 0) in the corresponding party.
%For this agent, the following differentiation is also presented: 
%
%\vskip 0.5 cm
%@@@@@@@@@@@@@@@{\bf (PER ALESSANDRO: CONTROLLARE  I SEGNI DI UGUALE NEL PROGRAMMA E  qui di seguito)}
%\vskip 0.5 cm
%
%- strong believer in Red party, with $1>IR_k>0.5$; 
%
%- strong believer in Blue party, with $1>IB_k>0.5$;
%
%- weak believer in Red party, with $0<IR_k<0.5$; 
%
%- weak believer in Blue party, with $0<IB_k<0.5$;
%
%\vskip 0.5 cm
%@@@@@@@@@@@@@@@@@@@@@@@@@@@@@@@@@@@@@@@@@@@@@@@@@@@@
%\vskip 0.5 cm
On the other hand, a given undecided agents $A_k$ (with $O_k=Y$) will start with both $IR_k$ and $IB_k$ randomly chosen in the interval $[I_{min},I_{max}]$.
\\
During the simulation ($t>0$), we assume that any agent $A_k$ belonging to one of the two parties can exert a different influence (social pressure) on his/her neighbors in order to induce them to change opinion or, in case of indecision, to assume an opinion. We reasonably define such a pressure as $P_k = int [IR_k * 10]$ or $P_k = int [IB_k * 10]$, depending on the status of the agent: it is therefore represented by an integer number included in the interval $[0,10]$. This means that strong believers of a party will have a greater influence on their neighbors, whereas weak believers will have only a little chance of convincing other people to change political preference or to assume a new one.  
\\
\\
\textit{Opinion dynamics rules}
\\
The dynamics of opinions (voting intentions) is very simple: we assume that, at each discrete time step $t$, an agent $A_k$ - with a given status $O_k$ - changes his/her preference or assumes a new one depending on both the opinion and influence at time $t-1$ of his/her neighbors $\{N_k\}$ who belong to a party and on his/her intensity of believing $IR_k$ or $IB_k$. More precisely, we define a vector $V_k(t-1)$ which contains the preferences $O_j(t-1)$ (i.e. the variables $R$ and $B$) of the neighbors $A_j \in \{N_k\}$ at time $t-1$ weighted by their influence. In other words, each preference $O_j$ will occur in the vector as many times as specified by the influence $I_j$ of the corresponding agent. Then, we randomly select one element $v$ of this vector and we change the values of $IR_k$ or $IB_k$ by distinguishing the following cases:
 
\begin{itemize}
\item[(i)] If $O_k=R$ and $v=R$, we increase the value of the variable $IR_k$ of a given quantity $\delta I$; if $O_k=R$ and $v=B$, we decrease the value of the variable $IR_k$ of the same quantity $\delta I$; if, after this operation, $IR_k>1$ or $IR_k<0$, we set, respectively, $IR_k=1$ or $IR_k=0$; if $IR_k=0$ we change the status of the agent $A_k$ from $O_k=R$ into the state $O_k=B$, then we assign to the variable $IB_k$ a new random value in the interval $[I_{min},I_{max}]$ and we put $IR_k=0$; 
\item[(ii)] If $O_k=B$ and $v=B$, we increase the value of the variable $IB_k$ of a given quantity $\delta I$; if $O_k=B$ and $v=R$, we decrease the value of the variable $IB_k$ of the same quantity $\delta I$; if, after this operation, $IB_k>1$ or $IB_k<0$, we set, respectively, $IB_k=1$ or $IB_k=0$; if $IB_k=0$ we change the status of the agent $A_k$ from $O_k=B$ into the state $O_k=R$, then we assign to the variable $IR_k$ a new random value in the interval $[I_{min},I_{max}]$ and we put $IB_k=0$; 
\item[(iii)] If $O_k=Y$ and $v=R$, we increase the value of the variable $IR_k$ of a given quantity $\delta I / 10$; if, after this operation, $IR_k \geq 1$, we change the status of the agent $A_k$ into the new state $O_k=R$, then we assign to the variable $IR_k$ a new random value in the interval $[I_{min},I_{max}]$ and we put $IB_k=0$; 
\item[(iv)] If $O_k=Y$ and $v=B$, we increase the value of the variable $IB_k$ of the same quantity $\delta I / 10$; if, after this operation, $IB_k \geq 1$, we change the status of the agent $A_k$ into the new state $O_k=B$, then we assign to the variable $IB_k$ a new random value in the interval $[I_{min},I_{max}]$ and we put $IR_k=0$; s
\end{itemize}

Notice that the opinions updating process is a  parallel one: at each time step $t$ all the agents update simultaneously their opinion depending on the opinions of all his/her neighbors at $t-1$ (therefore each agent can change his/her status only once during a given time step). Notice also that undecided agents, not belonging to any party, cannot influence other agents but can be induced by their neighbors to assume a position. This implies that their number will decrease in time while the election day is approaching.    
\\
At the beginning of each simulation run, once fixed the number $N$ of agents, one has to choose the initial conditions for the status distribution. One choice could be assigning the status (color) at random, with a uniform distribution, therefore obtaining more or less the same size ($\sim N/3$) for the three groups (the two parties, red and blue, and the undecided component, yellow). Another choice is to assign again the colors at random among the agents, but fixing independently the initial sizes $N_R$ and $N_B$ of the two parties through the sliders, then obtaining the following size for the undecided component: $N_Y = N - (N_R + N_B)$. In order to better control the initial advantage of a given party over the other one, we will usually prefer to adopt this second option. 
\\
\\
\textit{Surveys description}
\\
As already explained, our goal is to simulate how, starting from a given (biased) initial condition for the two parties, the opinion dynamics is affected by the results of a certain number $n_S$ of subsequent (biased) surveys $S_k$ ($k=1,2,3...,n_S$), which are {\it external} to the community ("canonical ensemble" situation) and whose behavior follows different scenarios. Once fixed a given party, say Blue party, and two thresholds, $\%min$ and $\%max$, the scenarios we consider are the following:

Scenario 1: Blue always prevailing in the surveys (with a percentage of preferences fluctuating between $\%min$ and $\%max$);

Scenario 2: Blue initially increasing (from $\%min$ to $\%max$), then decreasing (from $\%max$ to $\%min$), in both cases with small random fluctuations; 

Scenario 3: Blue always increasing, with small random fluctuations (from $\%min$ to $\%max$).

Of course, in correspondence of a given percentage $\%$  of preferences for Blue party, the score for Red party will be $100-\%$.
\\  
The surveys occur with a certain periodicity during the time interval $[0,T_E]$, being $T_E$ the total simulation time, coincident with the final "election day". Typically, we will fix $T_E=4000$ time steps, each one corresponding to one hour of real time: the global duration of the electoral campaign is therefore $4000h$, i.e. $166$ days, i.e. five and half months, approximately. The time interval among subsequent surveys is randomly chosen in the range $T_S - 24h, T_s + 24h$, centered around the value $T_S$ so that, on average, $n_S = T_E / T_S$. 
\\
After a survey $S_k$, the scores of the selected scenario will be stored in two variables $n_R(S_k)$ and $n_B(S_k)$. Furthermore, the "trend" evolution of these two preferences components are shown, i.e. the two differences: 
\\
\begin{equation}
\label{trend-differences}
\begin{split}
& d_R(S_k) = n_R(S_k) - n_R(S_{k-1}) \\
& d_B(S_k) = n_B(S_k) - n_B(S_{k-1})  
\end{split}
\end{equation}

The sign of each difference in eq.(\ref{trend-differences}) will indicate whether the trend for the corresponding component is positive or negative (and their value will indicate its strength). An initial survey at $t=0$ ($S_0$) is performed in order to show resulting trends since the first survey $S_1$ after the first $T_S$ time steps.
\\
The model allows to focus a fundamental point, which is the reaction of people to the survey's results (we assume that results are immediately advertised by mass media).
It is worth to notice that we assume that mass media, normally stress only information about scores and trends of the two parties, while the score of the undecided people, which in our model is surely decreasing in time, should be considered actually without any effect. 
\\
We also assume that the supporters of a given party will reinforce their opinion after a favorable survey, while the supporters of the survey-loosing party could either increase or decrease their convincement  depending on their personal characterization (i.e. if they are either strong or weak believers). On the other hand, also the undecided agents could increase their propensity to become believers depending on the score of the two parties.   
Thus, the reaction to a survey is modeled as it follows.
\\
After knowing the result of a survey $S_k$:
\\
- all the agents belonging to the party with the best score (the red party if $n_R(S_k) > n_B(S_k)$, the blue one in the opposite case) will increase their $IR_i$ or $IB_i$ of the quantity $\delta S$, i.e. they will reinforce their believing; tipically, $\delta S$ is greater than $\delta I$ and linearly increases with the difference between the scores of the two parties, i.e. $\delta S=\delta S_{min} + \beta_S \vert n_R(S_k) - n_B(S_k) \vert$;
\\
- at the same time, all the agents belonging to the loosing party, will increase their corresponding intensity ($IR_i$ or $IB_i$) of the same quantity $\delta S$ with probability $IR_i$ or $IB_i$, or will decrease their corresponding intensity with probability $1-IR_i$ or $1-IB_i$;
\\
- finally, all the undecided agents, will increase of the quantity $\delta S$ the intensity ($IR_i$ or $IB_i$) corresponding to the party with the best score.
\\   
An analogous procedure is then repeated, for all the agents and at the same time step, by comparing the trends $d_R(S_k)$ and $d_B(S_k)$ of the two parties: 
\\
- if $d_R(S_k)>0$ and $d_B(S_k)<0$, the red agents and the undecided agents will increase their $IR_i$ of the quantity $\delta S_{min}$ (i.e. in this case the increment does not depend on the difference between the scores of the two parties), while the blue agents either will increase their $IB_i$ of the quantity $\delta S_{min}$ with probability $IB_i$ or will decrease it with probability $1-IB_i$; 
\\
- if $d_B(S_k)>0$ and $d_R(S_k)<0$, the blue agents and the undecided agents will increase their $IB_i$ of the quantity $\delta S_{min}$, while the blue agents either will increase their $IR_i$ of the quantity $\delta S_{min}$ with probability $IR_i$ or will decrease it with probability $1-IR_i$; 
\\
In such a way, in response of each survey, the values of $IR_i$ or $IB_i$ of all the agents will slightly change and this could induce, at the next time step, a rearrangement of their status $O_i$.     
\\
At the end of each simulation (i.e. at $t=T_E$), all the agents are called to participate at final elections. Of course, undecided agents will not vote. But also the supporters of the two parties do not participate with certainty at the elections: actually, they will go to the polling station only with a probability, which is directly proportional to their conviction (i.e. to their influence, which is equal to $IR_i$ or $IB_i$). Therefore, in general, the final election scores for the two parties will be lower than the corresponding voting preferences at $t=T_E$, i.e. there will be always a given number of non voting people (abstainers) higher than the final number of undecided people.

\section{Results of the simulations}

In this section we  present the results of the  simulations for one single typical event.  We  show the expected time behavior of  opinion dynamics for a test community with a biased voting orientation, either in presence or in absence of surveys (biased in the opposite way), and, in the latter case, its dependence on the adopted surveys scenario. Immediately after, we present the results of multi-event simulations, in order to extract some statistical evidence about the influence of surveys on the elections result.

%%%%%%%%%  FIG. 2
\begin{figure}[t]
\begin{center}
\includegraphics[width=1.7in,angle=0]{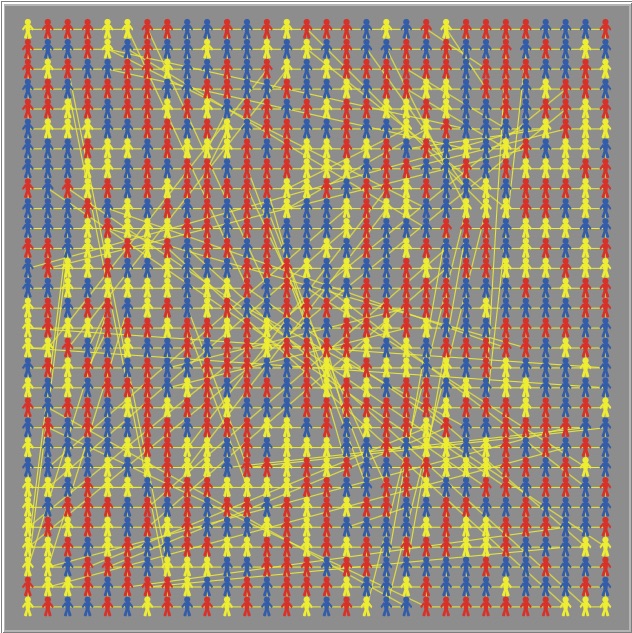}
\end{center}
\begin{center}
\includegraphics[width=1.51in,angle=0]{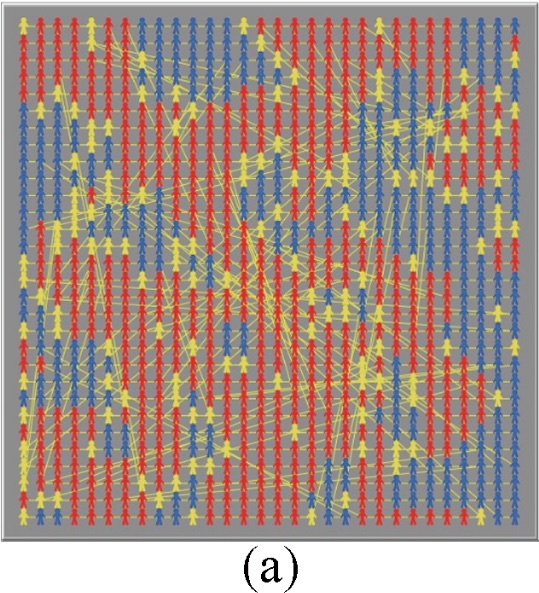}
\includegraphics[width=1.51in,angle=0]{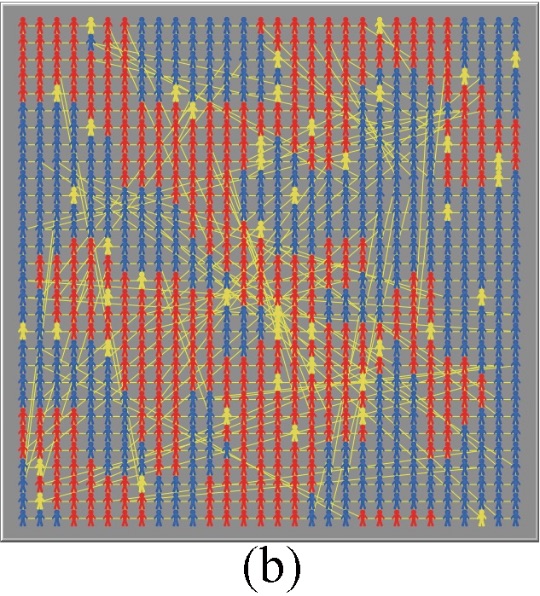}
\includegraphics[width=1.5in,angle=0]{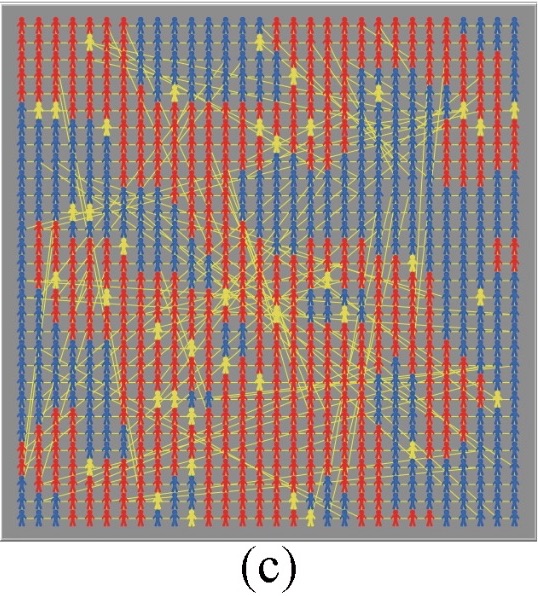}
\includegraphics[width=1.51in,angle=0]{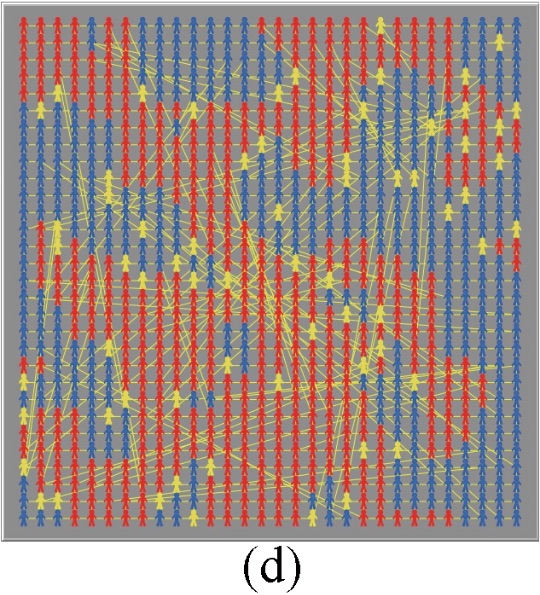}
\caption{
Top panel: initial configuration (at $t=0$) of our test community with $N=900$ agents. The different colors of the agents (red, blue or yellow) indicates, respectively, if a given individual, at the beginning of the simulation, is a believer of the Red party or of the Blue party or if the voter is still undecided. In this simulation setup the Red party always starts with a slight advantage ($40\%$) over the Bue party ($35\%$). Bottom panels: final configuration (at $t=T_E$) of the same community at the end of four simulations with different surveys scenarios: (a) No-surveys Scenario: Red party prevails; (b) Scenario 1: Blue party prevails; (c) Scenario 2: Red party prevails; (d) Scenario 3: Red party prevails.  
}
\label{fig2} 
\end{center}
\end{figure}
%%%%%%%%%%%%%%%%%%%%%%%%%%%%%%%%%%%%%%%%%%%%%%%%%%%%%%%%%%%%%%%%%%%%%%%%%%

\subsection{Single-event simulations}

We adopted the following setup for the control parameters for a typical simulation of one event:

- Number of agents: $N=900$;

- Fixed size, Red biased, initial conditions: $N_R=360$ ($40\%$), $N_B=315$ ($35\%$) and $N_Y=225$;

- Extreme values for the initial believing distribution: $I_{min}=0.10$ and $I_{max}=0.50$;

- Variation parameter for believing: $\delta I=0.002$;

- Minimum variation parameter for the reactions to the surveys: $\delta S_{min}=0.01$;

- Linear coefficient for the reactions to the surveys: $\beta_S=0.001$;

- Total simulation time, coincident with the "election day": $T_E=4000$ hours;

- Average time interval among subsequent surveys: $T_S=168$ hours (1 week);

\noindent
In order to test the typical effects of the three surveys' scenarios on the election results, it is useful to compare several single-event simulations realized by using {\it exactly the same topology} for the small-world community, reported in the top panel of Fig.\ref{fig2}, and starting from {\it exactly the same initial conditions} (at $t=0$), with a slight advantage of the Red party over the Blue party.\\

Before giving the details of each simulation, if we consider the bottom panels of Fig.\ref{fig2} and compare panel (a) - no-surveys scenario - with panels (b),(c) and (d), we immediately notice that the presence of surveys has three macroscopic effects on the final opinion distribution: (i) the reinforcement of the echo chambers, i.e. the accentuation of the spontaneous clustering of (red or blue) voting intentions which emerges among people due to the opinion dynamics; (ii) the consequent decrease of non-voting (yellow) people, who are normally located along the edges of opinion clusters; (iii) the ability to subvert, sometimes, the elections result (of course limited to the test community), as in the case of scenario 1 where, at variance with the no-surveys scenario, Blue party prevails.
%%%%%%%%%%%%%%%%%%%%%%%%%%%%%%%%%
%%%%%%%%%  FIG. 3
\begin{figure}
\begin{center}
\includegraphics[width=3.5in,angle=0]{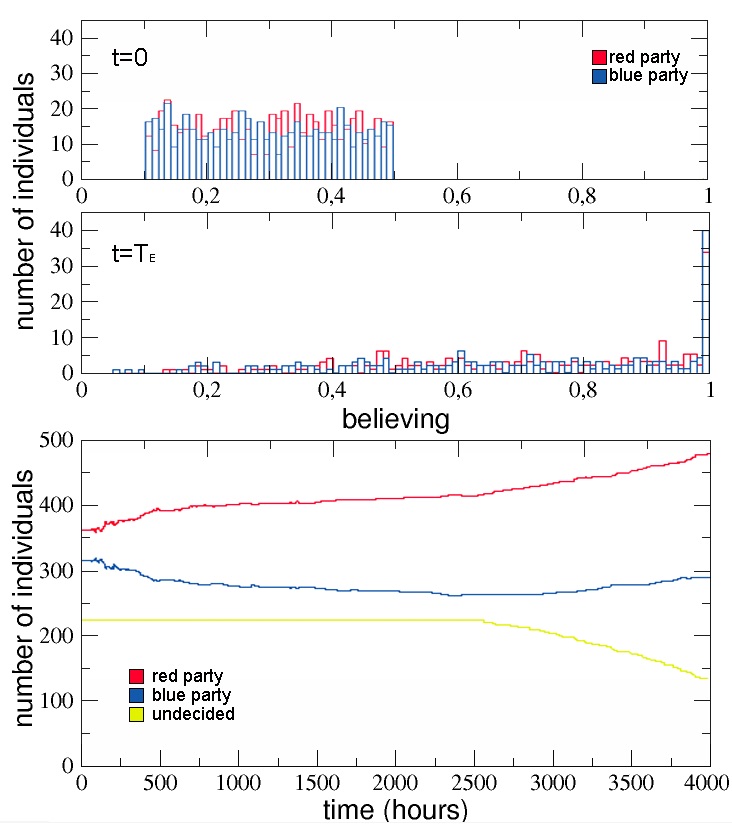}
\caption{{\it Single-event No-surveys Scenario.} Top panel: initial distribution (at $t=0$) of the believing intensities for the two parties. Middle panel: final distribution of believing intensities (at $t=T_E$). Bottom panel: time behavior of both the number of individuals voting for the two parties (red and blue) and the number of undecided (yellow). 
}
\label{fig3} 
\end{center}
\end{figure}
%%%%%%%%%%%%%%%%%%%%%%%%%%%%%%%%%%%%%%%
\\

%\vskip 0.5 cm
\noindent\textit{No-surveys Scenario}
\\
Let us start with the presentation of the time evolution of both believing and voting intentions in the no-surveys scenario. 

\noindent
In the top and the middle panels of Fig.\ref{fig3} we first show, respectively, the initial and final distributions of the believing in the two parties for the $900$ agents of the test community: starting at $t=0$ from their initial uniform values between $[0.1,0.5]$, the believing intensities for both the Red and the Blue parties evolve in time until, at the end of the simulation (i.e. at $t=T_E$) they reach a power-law like shape, with a pronounced peak in correspondence of their maximum allowed unitary value. In the bottom panel of Fig.\ref{fig3} it is shown the corresponding time behavior of the voting preferences for the three social components of the community: it clearly appears that, in absence of surveys, the initial slight numerical advantage of the Red party becomes stronger and stronger, in particular during the last two months when the undecided people, feeling the pressure of the incoming elections, start to assume a voting preference. Finally, at $t=T_E$ (elections day), the Red party component has reached  $N_R(T_E)=53\%$, against a $N_B(T_E)=32\%$ of the Blue party one and a $N_Y(T_E)=15\%$ of undecided. Then, following the rules of the model, the final distributions of believing translate into probabilities of going to the polling station for the Red and Blue components while, of course, the undecided component do not vote at all. Therefore, the elections result are the following: total percentage of voters $80.3\%$, score of the Red party $62.4\%$, score of the Blue party $37.6\%$. Comparing these values with the final composition of the voting preferences, we observe that the percentage of abstainers ($19.7\%$) is higher than that of the undecided component ($15\%$): this means that some of the weak believers of the two parties did not go to the polling station. However, Red party still maintain its advantage and prevails at the elections.               
%%%%%%%%%%%%%%%%%%%%%%%%%%%%%%%%%
%%%%%%%%%  FIG. 4
\begin{figure}
\begin{center}
\includegraphics[width=6.7in,angle=0]{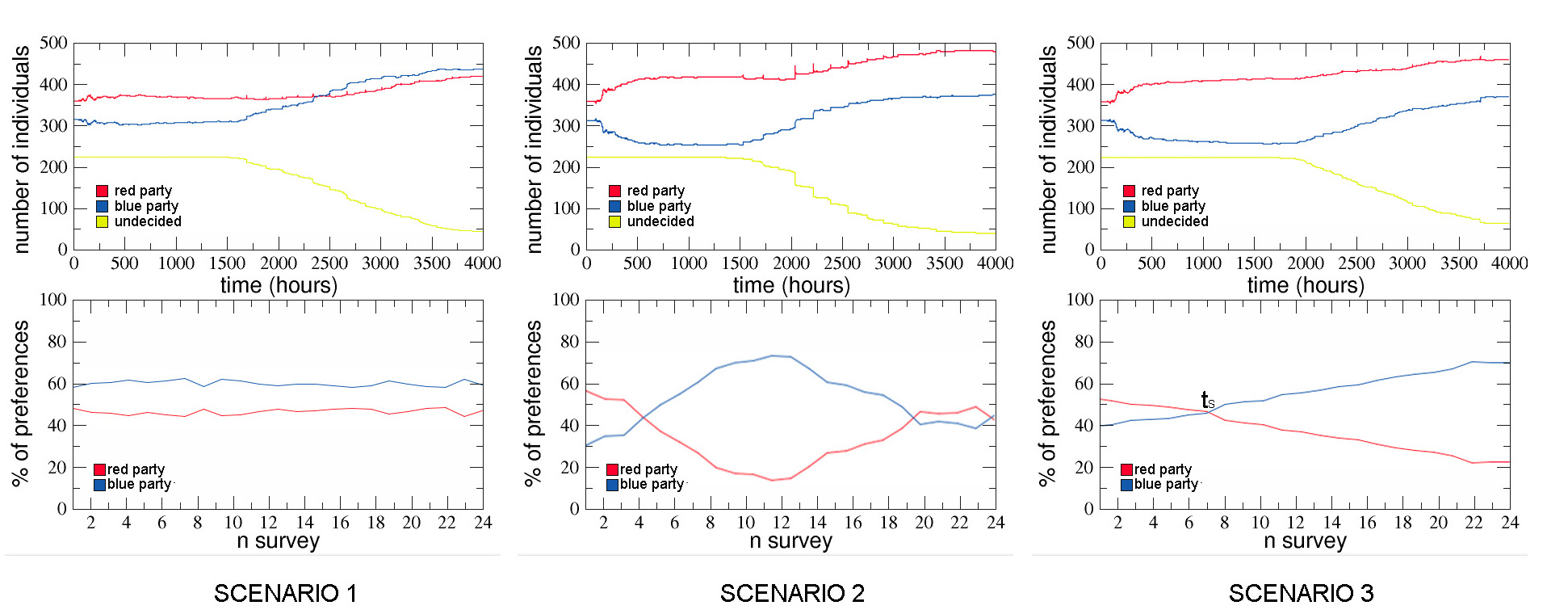}
\caption{ {\it Single-event Surveys Scenarios.}  Top panels: the time behavior of both the number of individuals voting for the two parties (red and blue) and the number of undecided (yellow) is shown for the three Surveys Scenarios considered. Bottom panels: the scores of the two parties for each one of the $n_S=24$ Surveys (one per week, in average) is reported for the three Scenarios considered.      
}
\label{fig4} 
\end{center}
\end{figure}
%%%%%%%%%%%%%%%%%%%%%%%%%%%%%%%%%%%%%%%
\\

%\vskip 0.5 cm

\noindent\textit{Surveys Scenarios}
\\
Let us consider, now, the three scenarios with surveys. 
\\
\noindent
In the top panels of Fig.\ref{fig4}, the time evolution of the voting preferences for the three components of the test community is reported for each one of the three surveys scenarios. In the bottom panels of the same figure, the corresponding scores of the two parties within each surveys scenario is reported for comparison. It clearly appears that only in Scenario 1, due to effect of the biased surveys score of the Blue party which is always over that one of the Red party (fluctuating between $54\%$ and $58\%$), the Blue party is able to recover its initial disadvantage and to prevails at the elections: in this case, the percentage of voting people is $90.9\%$ - $48.9\%$ for Red party, $51.1\%$ for Blue party - and that of abstainers is $9.1\%$. In the other two scenarios, the effects of the biased surveys are not relevant and the Red party remains always prevailing, like in the no-surveys case: in Scenario 2 (where the surveys score of the Blue party starts at $30\%$, initially goes up until $70\%$, then goes down again towards $40\%$) the percentage of voting people is $93.1\%$ - $56.8\%$ for Red party, $43.2\%$ for Blue party -, while in Scenario 3 (where the surveys score of the Blue party starts at $40\%$, overtakes that of the Red party at the surpass time $t_S$ and slowly goes up until $70\%$) the percentage of voting people is $88.9\%$ - $55.4\%$ for Red party, $44.6\%$ for Blue party. 

In conclusion, as already observed, even if sometimes it is not enough to overturn the initial bias (of $5\%$) in favor of the Red party, the introduction of surveys has always the effect of reducing the percentage of abstainers (by increasing the average believing of people) and of reinforcing the clustering of preferences into separated echo-chambers. On the other hand, it seems that a constant advantage in the surveys score (like in Scenario 1, where an average advantage of $56-44=12\%$ has been considered) is strictly necessary to the Blue party for having a chance to prevail at the elections in a community initially biased in favor of the Red party. But, how the magnitude of such an advantage does affect the elections results? And how these results are also affected by the range of variation of the surveys scores for the Blue party in Scenarios 2 and 3? In order to answer to these questions, and to obtain results more significant from a statistical point of view, in the next section we will perform a parametric analysis of the model by means of systematic multi-event simulations.

%%%%%%%%%%%%%%%%%%%%%%%%%%%%%%%%%
%%%%%%%%%  FIG. 5
\begin{figure}
\begin{center}
\includegraphics[width=3in,angle=0]{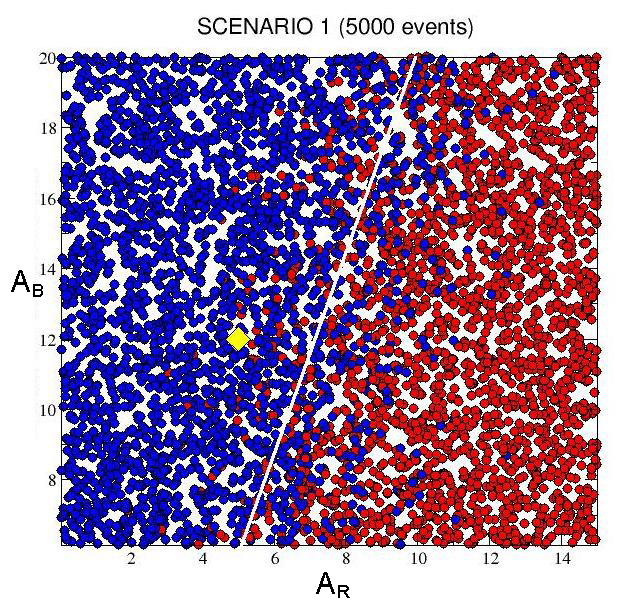}
\includegraphics[width=2.6in,angle=0]{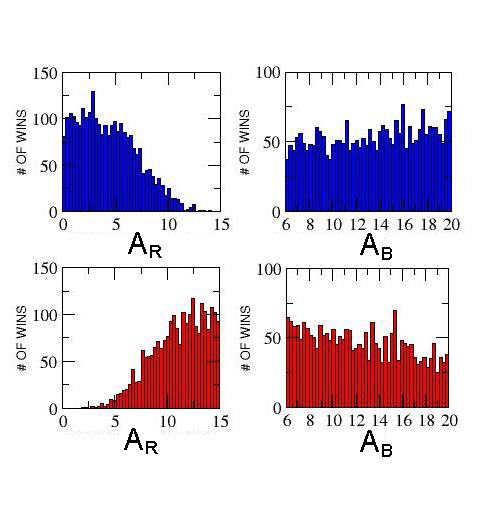}
\caption{{\it Multi-event Surveys Scenario 1}. In the diagram on the left are reported $5000$ points, each one corresponding to a single-event simulation, colored in blue or in red depending on the party prevailing at the elections, as function of both the initial biased advantage $A_R$ of the Red party in the percentage of believers (x-axis) and the biased average advantage $A_B$ of the Blue party in the surveys score (y-axis). In the four panels on the right, the number of wins for the Blue party (upper panels) and the Red party (bottom panels) are also reported as function of $A_R$ (left column) and $A_B$ (right column).           
}
\label{fig5} 
\end{center}
\end{figure}
%%%%%%%%%%%%%%%%%%%%%%%%%%%%%%%%%%%%%%%

\subsection{Multi-event simulations}

Let us start by exploring how the probability for the Blue party of overturning the initial disadvantage in terms of preferences with respect to the Red party in the context of Scenario 1 does depend on both the initial biased advantage in the percentage of believers of the Red party ($A_R$) and on the biased average advantage of the Blue party in the surveys score ($A_B$). 
  
In the left diagram of Fig.\ref{fig5}, the results of $5000$ different single-event simulations (with the same setup of the control parameters described in the previous section) are reported as colored points uniformly distributed as function of these two quantities ($A_R$ in the x-axis and $A_B$ in the y-axis): the color of each point indicates the party that won the elections in the corresponding event. In the four panels on the right, the number of wins for the Blue party and the Red party are reported separately, as function of $A_R$ (left column) and $A_B$ (right column) respectively. It is evident that, as one could expects, increasing the initial believing advantage $A_R$ of Red party (from $1\%$ to $15\%$) makes more difficult for the Blue party to overtake it and to prevail at the elections, while increasing its average advantage $A_B$ in the survey score (from $6\%$ to $20\%$) makes the victory of Blue party easier (in any case, if $A_R<2\%$ the Blue party always prevails at the elections, no matter its survey advantage). The inclined white line reported in the diagram helps the eye to appreciate this effect: to the left of this line the chance of prevailing of the Blue party is greater than that of the Red party, to the right the opposite holds. Notice that the single-event result where the Blue party won the elections in the survey Scenario 1 with  $A_R=5\%$ and $A_B=12\%$, discussed in the previous section and shown in Fig.\ref{fig3}, is consistent with this picture (the point $A_R=5\%$ and $A_B=12\%$ is indicated with a yellow diamond in the left diagram of Fig.\ref{fig4}).

%%%%%%%%%%%%%%%%%%%%%%%%%%%%%%%%%
%%%%%%%%%  FIG. 6
\begin{figure}
\begin{center}
\includegraphics[width=5in,angle=0]{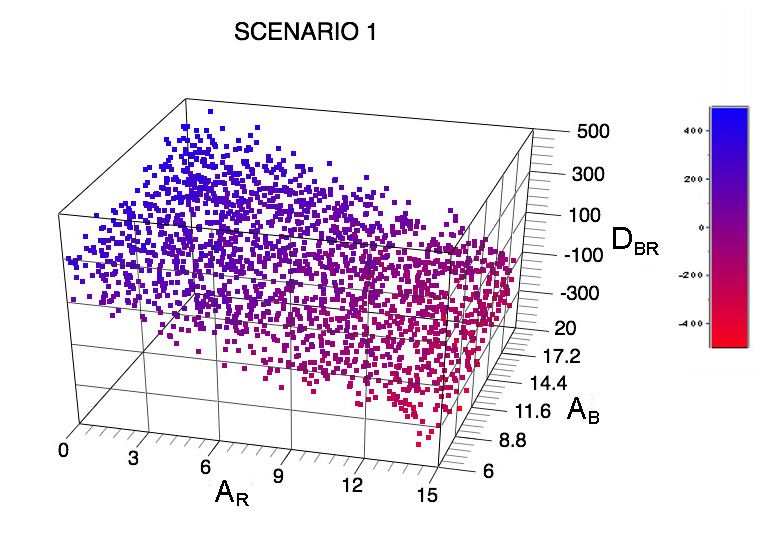}
\caption{ {\it Multi-event Surveys Scenario 1}. In this 3D diagram the same $5000$ single-event points of the previous figure are plotted as function of $A_R$, $A_B$ and $D_{BR}$. The latter, reported in the z-axis, is the difference between the number of votes taken by the Blue party at the elections and those taken by the Red party, and it is here used as a proxy of the probability of prevailing of the Blue party. The shades of color from Red to Blue helps to better appreciate the position of the points along the z-axis.   
}
\label{fig6} 
\end{center}
\end{figure}
%%%%%%%%%%%%%%%%%%%%%%%%%%%%%%%%%%%%%%%

In order to visualize in a different way the variation in the chance of prevailing of the Blue party as function of $A_R$ and $A_B$, in Fig.\ref{fig6} we add a third z-axis to the diagram of Fig.\ref{fig5}: in this axis the difference $D_{BR}$ between the number of votes taken by the Blue party at the elections and those taken by the Red party is reported as a proxy of the probability of prevailing of the Blue party. A color scale for the z variable, going from Red (for $D_{BR}<0$) to Blue (for $D_{BR}>0$), applied to the single-event points, helps the eye to appreciate both the decrease of that probability along the $A_R$ axis and the increase along the $A_B$ axis.    

Let us now to go to the multi-event analysis of the other two surveys scenarios, starting with Scenario 2. 

As already shown in Fig.\ref{fig4}, Scenario 2 provides that the surveys score of the Blue party starts at $30\%$, initially goes up until $70\%$, then goes down again. In this scenario, the surveys give the Blue party as leading party only in the central part of the single-event simulation, while the Red party results to be in advantage at the beginning and at the end of the time period considered. Once fixed this range of variation for the surveys score, and adopting again the same setup of parameters of the previous section, we perform a multi-event simulation with $2000$ events, each one with a different value for the initial advantage $A_R$ of the Red party but leaving it to be also negative ($-10<A_R<10$): this means that now the initial bias in the composition of believing of the test community can also be in favor of the Blue party (when $A_R<0$).     

%%%%%%%%%%%%%%%%%%%%%%%%%%%%%%%%%
%%%%%%%%%  FIG. 7
\begin{figure}
\begin{center}
\includegraphics[width=4.5in,angle=0]{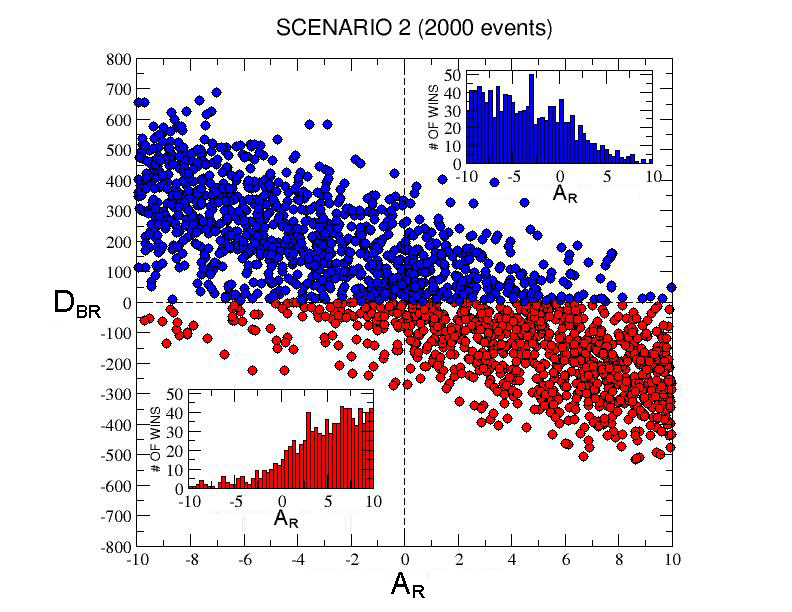}
\caption{{\it Multi-event Surveys Scenario 2}. In the diagram are reported $2000$ points, each one corresponding to a single-event simulation, colored in blue or in red depending on the party prevailing at the elections, as function of both the initial biased advantage $A_R$ of the Red party in the percentage of believers (x-axis) and the difference $D_{BR}$ between the number of votes taken by the Blue party at the elections and those taken by the Red party (y-axis). In the insets, the number of wins for the Blue party (upper panel) and the Red party (bottom panel) are also reported as function of $A_R$.}
\label{fig7} 
\end{center}
\end{figure}
%%%%%%%%%%%%%%%%%%%%%%%%%%%%%%%%%%%%%%%

In the diagram shown in Fig.\ref{fig7}, we report the points corresponding to the $2000$ single-event simulations of Scenario 2 as function of the $A_R$ value (x-axis) and also of the difference $D_{BR}$ between the number of votes taken by the Blue party and those taken by the Red party at the elections (y-axis). As usual, the points are colored in blue or in red depending on the party prevailing at the elections. Of course, in this case all the points above the x-axis (i.e. with $D_{BR}>0$) will be blue while those below the x-axis will be red. The distribution of the points in the diagram clearly indicates that the elections winner strictly depends on the initial advantage in the number of believers: by increasing $A_R$, such an advantage will be more and more consolidated by the opinion dynamics during the simulations and the initially favored party shall prevail at the elections with higher probability. 

%%%%%%%%%%%%%%%%%%%%%%%%%%%%%%%%%
%%%%%%%%%  FIG. 8
\begin{figure}
\begin{center}
\includegraphics[width=4.5in,angle=0]{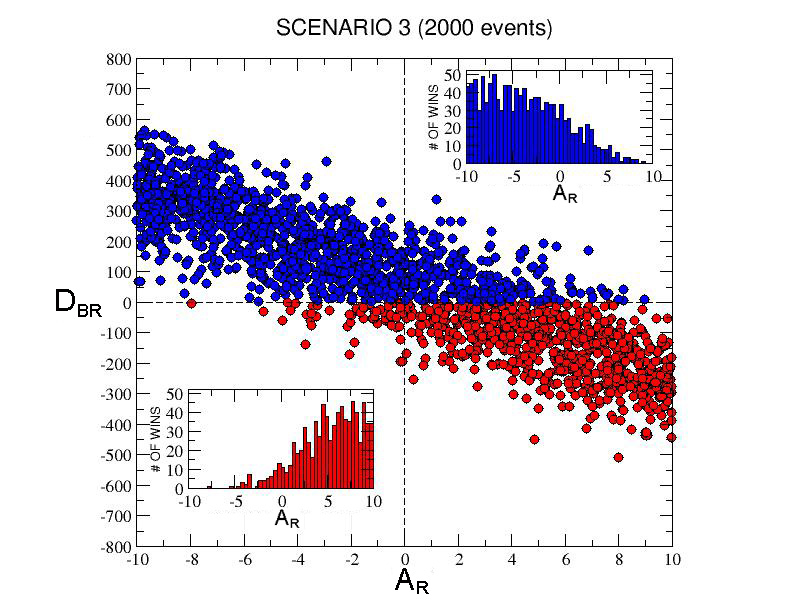}
\caption{{\it Multi-event Surveys Scenario 3}. In the diagram are reported $2000$ points, each one corresponding to a single-event simulation, colored in blue or in red depending on the party prevailing at the elections, as function of both the initial biased advantage $A_R$ of the Red party in the percentage of believers (x-axis) and the difference $D_{BR}$ between the number of votes taken by the Blue party at the elections and those taken by the Red party (y-axis). In the insets, the number of wins for the Blue party (upper panel) and the Red party (bottom panel) are also reported as function of $A_R$.}

\label{fig8} 
\end{center}
\end{figure}
%%%%%%%%%%%%%%%%%%%%%%%%%%%%%%%%%%%%%%%

Anyway, the effects of the surveys is slightly asymmetric for the two parties: actually, from the details of the histograms shown in the two insets, where the number of wins as function of $A_R$ is reported for the two parties, it results that the $23.5\%$ of the Blue party wins (over a total of $1090$) occur when it starts from disadvantageous initial conditions, i.e. when $A_R>0$, against $11.3\%$ of the Red party wins (over a total of $910$) for $A_R<0$. This means that it seems more convenient for a party to be favored by the surveys scores (with an increasing trend) in the first half of the election campaign rather than in the second part, provided that the duration of its central leading time period is quite wide. Notice also that, for a given value of $A_R$, the spreading of $D_{BR}$ along the y-axis is quite large, i.e. the gap of votes between the winning and the losing party can assume many different values, just linked to the duration of the central leading time period.                         

An analogous effect appears also in the case of the Scenario 3 (again with the same setup of the previous section), where the surveys score of the Blue party starts at $40\%$, but immediately increases and, after overtaking the Red party at a given time $t_S$, slowly goes up until $70\%$ (see Fig.\ref{fig4}). In fact, the results of the multi-event analysis reported in Fig.\ref{fig8}, consisting again of $2000$ events, show that the $21.2\%$ of the Blue party wins (over a total of $1197$, more than those of Scenario 2) occur when it starts from disadvantageous initial conditions ($A_R>0$), against only $5.5\%$ of the Red party wins (over a total of $803$) for $A_R<0$: this implies that when the Blue party starts with adverse surveys but then surpasses the Red party showing a constant positive survey trend, its total number of wins increases with respect to the Scenario 2, in particular (of course) when $A_R<0$ (the number of wins when $A_R>0$ remains more or less the same as in the Scenario 2). Furthermore, the spreading of $D_{BR}$ along the y-axis for each value of $A_R$ is reduced with respect to the previous scenario. 

\begin{figure}
\begin{center}
\includegraphics[width=6.2in,angle=0]{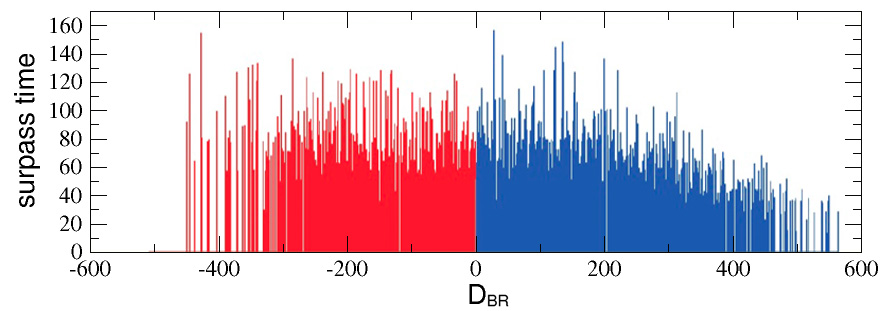}
\caption{{\it Multi-event Surveys Scenario 3}. Behavior of the surpass time $t_S$ as function of the election results difference $D_{BR}$ for both the Red ($D_{BR}<0$) and Blue ($D_{BR}>0$) parties. }
\label{fig9} 
\end{center}
\end{figure}

In this scenario it is also interesting to see how the surpass time $t_S$ influences the election result. In Fig.\ref{fig9}, where the surpass time is plotted as function of the usual difference $D_{BR}$ between the number of votes taken by the Blue party and those taken by the Red party, we may notice that the value of the average surpass time $<t_S>$ is lower when the Blue party prevails at the elections ($51.6$ against $61.2$): this means that the probability for the Blue party of overturning the initial disadvantage and winning the electoral competition within is higher when the surpass in the surveys happens quite soon. In particular, the gap $D_{BR}>0$ between the votes of Blue party and those of the Red party is maximum when $t_S < 30 $ $days$, i.e. when the surpass falls within the first month (simmetrically, when $t_S$ falls near the election day $T_E$, the gap is also quite high but in favor of the Red party - i.e. $D_{BR}<0$).   

%%%%%%%%%%%%%%%%%%%%%%%%%%%%%%%%%
%%%%%%%%%  FIG. 9
%%%%%%%%%%%%%%%%%%%%%%%%%%%%%%%%%%%%%%%

\section{Conclusions}

 We have presented  an agent-based model on a small-world realistic topology that should be able to  capture  the effective influence  of surveys in orienting the opinions of voters before elections. The dynamics of electoral consensus was investigated by considering different scenarios with two coalitions/parties and a third group of undecided voters (but, of course, the model could be be easily extended to more than two parties). We have shown that the effect of periodic public surveys on the opinions of a relatively small community of agents, if compared with an identical situation but without surveys, is twofold: on one hand, surveys do reinforce the so called 'echo chambers', i.e. accentuate the spontaneous clustering of voting intentions  emerging among people due to the opinion dynamics; on the other hand, they can change the final electoral result (of course limited to the considered community) and let the party, that otherwise would lose, to win the electoral competition at the end of the examined period. Of course, the proposed theoretical model would need to be supported with real data in order to calibrate the internal  parameters and become possibly a reliable predictive model after the experimental validation of the initial assumptions. In this respect, the surveys' scenarios considered in this paper are just examples to show the effectiveness of this methodology. In fact, we think that this kind of analysis could suggest possible strategies in order to manage and investigate in detail the formation of electoral consensus in political competitions. In practice, by knowing the voting intentions of a community at the beginning of the trial period and the external survey scenario during a large portion of the same period, our model can possibly infer (with a certain probability) the final distribution of votes in the community. 

\vskip 0.5 cm 
 
 \section*{Acknowledgments}
 This study was partially supported by the FIR Research Project 2014 N.ABDD94 of the University of Catania, ITALY.
  
\vskip 0.5 cm 
  
%\bibliographystyle{abbrv} 
%\bibliography{ABiondo_Survey}

\end{document}